\documentclass{acm_proc_article-sp}
\usepackage{times}
\usepackage{helvet}
\usepackage{courier}
\usepackage{graphicx}
\usepackage{url}

\usepackage{lipsum}

\begin{document}

\title{Echo chamber amplification and disagreement effects in the political activity of Twitter users}
\numberofauthors{2}
\author{
%
\alignauthor
Kirill Dyagilev\\
       \affaddr{Department of Electrical Engineering}\\
       \affaddr{Technion - Israel Institute of Technology}\\
       \email{kirilld@tx.technion.ac.il}
\alignauthor
Elad Yom-Tov\\
       \affaddr{Microsoft Research}\\
       \affaddr{Herzliya, Israel}\\
       \email{eladyt@microsoft.com}
}

\date{16 June 2014}

\maketitle
\begin{abstract}
Online social networks have emerged as a significant platform for political discourse.
In this paper we investigate what affects the level of participation of users in the political discussion.
Specifically, are users more likely to be active when they are surrounded by like-minded individuals, or,
alternatively, when their environment is heterogeneous, and so their messages might be carried to people with differing views.

To answer this question, we analyzed the activity of approximately $200,000$ twitter users who expressed
explicit support for one of the presidential candidates of the $2012$ US presidential election.
We quantified the level of political activity (PA) of users by the fraction of political tweets in their posts,
and analyzed the relationship between PA and measures of the users' political environment.
These measures were designed to assess the likemindedness, e.g., the fraction of users with similar political views,
of their virtual environment and, for a subset of approximately $20,000$ twitter users, of their geographic environment.

Our results showed that the highest levels of PA are usually obtained
by users in politically balanced virtual environment.
This is in line with the disagreement theory of political science that states that a user’s
political activity is invigorated by the disagreement with their peers.
Our results also show that users surrounded by politically like-minded
virtual peers tend to have a low level of PA.
This observation contradicts the echo chamber amplification theory
that states that a person tends to be more politically active when
surrounded by like-minded people.
Finally, we observe that the likemindedness of the geographical environment
does not have a significant effect on the level of PA of users.

We thus conclude that the level of political activity of the Twitter users is independent of
the likemindedness of their geographical environment and is correlated with
likemindedness of their virtual environment. The exact form of correlation manifests
the phenomenon of disagreement and, in a majority of settings, contradicts the echo chamber amplification theory.
\end{abstract}




\section{Introduction}
In the recent years, online social networks have emerged
as a significant platform for discussion and dissemination of political information.
For example, $2011$ Pew surveys \cite{lot:pew_2011} found that 22\% of adult Internet users participated in
political campaigns through at least one of the major social media platforms (Twitter, Facebook, Myspace)
during the 2010 US elections. Similarly, it was found in \cite{lot:pew_2012} that, in year 2012,
34\% of social network users posted their own thoughts on political or social issues,
and 38\% of users ``liked" and reposted political posts of others.

This increasing importance of social media and the relative convenience of its analysis
attracted attention from academic researchers.
Among the questions that have been investigated are: can future election results be predicted (e.g., \cite{lit:gayo_limits}),
is political information on Twitter credible \cite{lit:castillo_inf_cred,lit:morris_tweeting_believing,lit:ratkiewicz_polit_abuse},
who are users whose opinion on a certain subject is influential \cite{lit:barbieri_topic_aware_infl,lit:weng_twitterrank},
and how to leverage anonymized web search queries to analyze and visualize political issues \cite{lit:weber_webQ_polit}.

In this paper we analyze social factors associated with the level of participation of
users in the political discussion. Two complementing theories were suggested in the scientific literature
to explain the interaction between likemindedness of ones' social environment and of the level
of political activity:
\begin{itemize}
\item\textbf{Echo chamber and echo chamber amplification.} In \cite{lit:echo_chamber}, the author shows that people tend
to look for cognitive comfort by discussing their opinions with like-minded people.
Their opinions are thus echoed and reinforced by their social peers; creating an \emph{echo chamber} effect.
In the context of the web, the echo chamber effect is achieved when people follow blogs and news sources
that do not challenge their political opinions. This theory predicts that people
in comforting environments such as echo chambers will exhibit an \emph{amplified} level of political activity.
\item\textbf{Disagreement effect.} A large body of political-science literature
\cite{lit:moy_predicting,lit:mutz2006hearing,lit:nir_disagreement,lit:pattie_conversation}
explores the effect that \emph{disagreement}, i.e., having political opinions different from your (non-virtual) social peers,
has on your political activity.
Nir \cite{lit:nir_disagreement} shows that disagreement has a dual effect.
A politically isolated person, in the sense that \emph{all} of their peers disagree with their opinion,
tends to exhibit a lower than average level of political activity.
However, a person with politically heterogeneous social peers tends to exhibit a higher level of political activity
(see also \cite{lit:pattie_conversation}), even when compared to people completely surrounded with like-minded peers
(echo chamber). This theory predicts that the level of political activities should be attenuated
for isolated users and have a dominant peak for heterogeneous social environments.
\end{itemize}
In this paper we test the presence and the relative importance of these two phenomena
in political discussions in Twitter.
We also test a conjecture that likemindedness of both the virtual (web) environment
and the physical (geographical) environment have effect on a user's level of political activity.

\section{Methods}\label{sec:data_set_desc}
In this paper we analyze data from Twitter - a micro-blogging service that allows users to post
short ``tweets" and to receive tweets made by other users by ``following" their Twitter feeds, thus
creating a social network of Twitter accounts. Our data extraction technique largely follows
methods from our previous work in \cite{lit:my_msr_journal}.

In our analysis we extract and analyze four overlapping sets of Twitter users.
We begin by extracting \emph{Raw-DS}, which is a large set of users with known political affiliations
and known levels of political publishing activity before 2012 US presidential election.
We then extract a \emph{Retweet-network} graph that is used as an approximation of a social network
connecting Twitters users in Raw-DS.
Using connections defined by Retweet-network, we extract two subsets of Raw-DS: \emph{Follower-DS}
and \emph{Followed-DS}. These sets of users are used to analyze the relation between
users' level of political activity and the likemindedness of their neighbors in the Retweet-network.
Finally, we extract a small subset \emph{Geographical-DS}
of users of Raw-DS that have enough available geo-spatial information to identify counties they reside in.
The Geographical-DS data set is then used to analyze the relation between users' level of political activity and
the likemindedness of their physical environment.
We proceed formally define each one of these sets.\\
$ $\\
\textbf{Raw-DS.}
We begin by extracting a large set of users with known political affiliations and
known levels of political publishing activity before 2012 US presidential election.
We denote this set of users by \emph{Raw-DS}.
In what follows, we alternatively refer to these users as pro-Obama and pro-Romney, respectively.

To this end, we employed the method described in \cite{lit:my_msr_journal}.
Namely, we looked for specific highly-partisan hashtags (a single word preceded by ``\#" sign,
listed in Table \ref{Table:hashtags_identPolitUsers}) among the tweets made in the 10 days following Election Day.
We picked this method over other existing methods (e.g., \cite{lit:conover_predicting_politAlign,lit:twit_voting_beh_from_lingExp,lit:pennacchiotti_demRepAndStarbucks})
because of ease of its implementation and accuracy (above $95\%$) that is higher than in other solutions.
The simplicity and the higher accuracy of our method come at the expense of a smaller recall than that of
other existing methods. However, the obtained user population was large enough for meaningful analysis.

As in \cite{lit:my_msr_journal}, we found a total of $372,769$ pro-Obama users and
$22,902$ pro-Romney users.

\begin{center}
\begin{table}[ht]
\hfill{}
\begin{tabular}{| c | l |}
  \hline
  $\quad$Affiliation$\quad$ & $\quad$Used hashtags$\quad$\\
  \hline
  Pro-Obama & \#voteobama, \#obama2012, \#goobama,\\
  & \#obamabiden, \#guardthechange, \\
  & \#4moreyears,\\
  & \#forward, \#forwardwithobama,\\
   & \#obamaforpresident,\#igoobama\\
  \hline
  Pro-Romney & \#romenyryan2012, \#voteromneyryan,\\
   &\#voteromney, \#benghazi, \#nobama, \\
   &\#imwithmitt, \#americascomebackteam, \\
   &\#fireobama, \#teamprolife, \#gogop\\
  \hline
\end{tabular}
\hfill{}
\caption{List of hashtags used for identification of the political affiliation of users.}\label{Table:hashtags_identPolitUsers}
\end{table}
\end{center}

We then extracted all tweets published by users in Raw-DS during the three and a half months period between Aug. 1st and Nov. 15th $2012$.
This interval includes tweets published, roughly, three months before the Election Day (Nov. 6th) and
ten days after it.

Overall, there were $55,740,001$ tweets. As in \cite{lit:my_msr_journal}, we used hashtags such as
"\#election2012" (see Table \ref{Table:hashtags_identPolitTweets} for the complete list) to extract a subset
of $465,842$ tweets on political issues.
\begin{center}
\begin{table}[ht]
\hfill{}
\begin{tabular}{| l |}
  \hline
  List of hashtags:\\
  \hline
  \tiny{$ $}\\
  All hashtags from Table \ref{Table:hashtags_identPolitUsers}, \#tcot , \#election2012, \#gop,\\
  \#romney,\#obama, \#elections, \#president\\
  \hline
\end{tabular}
\hfill{}
\caption{List of hashtags used to identify political tweets.}\label{Table:hashtags_identPolitTweets}
\end{table}
\end{center}
Given the number of political and non-political tweets made by each user in Raw-DS, we were able to
calculate their political activity (\emph{PA}), which we quantified
as the fraction of political tweets among their posts.\\
$ $\\
\textbf{Retweet-network.}
We next inferred the edges of the social graph that connects users in Raw-DS.
There are several commonly used proxies for the social connections between Twitter users.
For instance, one approach is to assume that there exits a directed edge from user A to user B if user A follows user B's Twitter feed.
Another approach is to define an edge from user A to user B if user A ``retweeted" one of user B's posts.
We choose the latter approach as it indicates a stronger connection between users. Namely,
user A is more than just skimming through user B's Twitter feed, user A also actively engages with its content.
We refer to corresponding social network over users in Raw-DS as the \emph{Retweet-network}.
We keep the terms ``follower" and ``followed" to describe the relationship between users in the Retweet-network.

The Retweet-network contains $201,362$ edges for $392,995$ nodes, which implies that the network is highly sparse.
Table \ref{Table:RTnetwork_stats} shows the number of edges between users for the four possible pairs of political affiliations.
\begin{center}
\begin{table}
\hfill{}
\begin{tabular}{| c | c | c |}
  \hline
  $\space$ User pair $\space$  &   $\space$  Number of edges$\space$\\
   $\space$ &   $\space$  (in thousands) $\space$\\
  \hline
  (pO,pO) & $655$\\
  (pO,pR) & $51$\\
  (pR,pO) & $48$\\
  (pR,pR) & $42$\\
  \hline
\end{tabular}
\hfill{}
\caption{Connectivity statistics in Retweet-network. Acronym pO stands for pro-Obama user and acronym pR stands for pro-Romney user.
The first user in the pair follows the second user in the pair, e.g., in a pair (pO,pR) a pro-Obama user follows a pro-Romney user.}\label{Table:RTnetwork_stats}
\end{table}
\end{center}
We note a large number of retweets that cross party lines,
i.e., a tweet made by a pro-Obama user is retweeted by a pro-Romney user or vice versa.
In \cite{lit:my_msr_journal}, we show that some of these retweets are part of a political debate.
In particular, when a link to a political article published by a pro-Obama user is retweeted by a pro-Romney user,
the text accompanying the retweeted link is likely to be modified, usually
to interpret the link according to the users' own point of view \cite{lit:my_msr_journal}.\\
$ $\\
\textbf{Follower-DS.} Using Retweet-network, we extracted a subset\\ \emph{Follower-DS} of users in Raw-DS
that have at least one follower. This subset contains $3,831$ pro-Romney users and $48,858$ pro-Obama users.
For users in Follower-DS, we refer to the likemindedness of their followers as
\emph{follower-LM} and quantify it as the fraction of users' followers that share their choice of candidate.\\
$ $\\
\textbf{Followed-DS.} Similarly to Follower-DS, we extracted a subset\\ \emph{Followed-DS} of users in Raw-DS
that follow at least one user. This subset contains $10,217$ pro-Romney users and $187,374$ pro-Obama users.
For users in Followed-DS, we refer to the likemindedness of Twitter feeds they follow as
\emph{followed-LM} and quantify it as the fraction of Twitter feeds followed by these users
that share their choice of candidate.
We note that this separation between Follower-DS and Followed-DS is meaningful
since only $12.5$\% of edges in Retweet-network are reciprocated.\\
$ $\\
\textbf{Geographical-DS.}
Finally, we identified a subset \emph{Geographical-DS} of users in Raw-DS
with enough geo-spatial information to identify counties they reside in.
To this end, we began by extracting a larger subset of users that
provided their geographical location (in terms of GPS coordinates) in at least two of their tweets.
For each such user, we calculated their average location by taking the mean value of GPS coordinates.
In order for this average location to be representative,
we discarded all users with the maximal distance between user's locations greater than $50$ kilometers.
We further discarded all users with the average location outside of the United States.
The remaining subset Geographical-DS contains a total of $1,083$ pro-Romney users and $18,475$ pro-Obama users.
For each user in Geographical-DS we use their average location to identify the county this user resides in
and obtain the official voting record for this county.
Given this information we are able to calculate the likemindedness of user's geographical environment (\emph{geographical-LM}),
which is quantified as voting share of user's candidate in their county.

\section{Results}
\begin{figure}
\centering
\begin{minipage}{0.4\textwidth}
\centering
\includegraphics[scale=0.4]{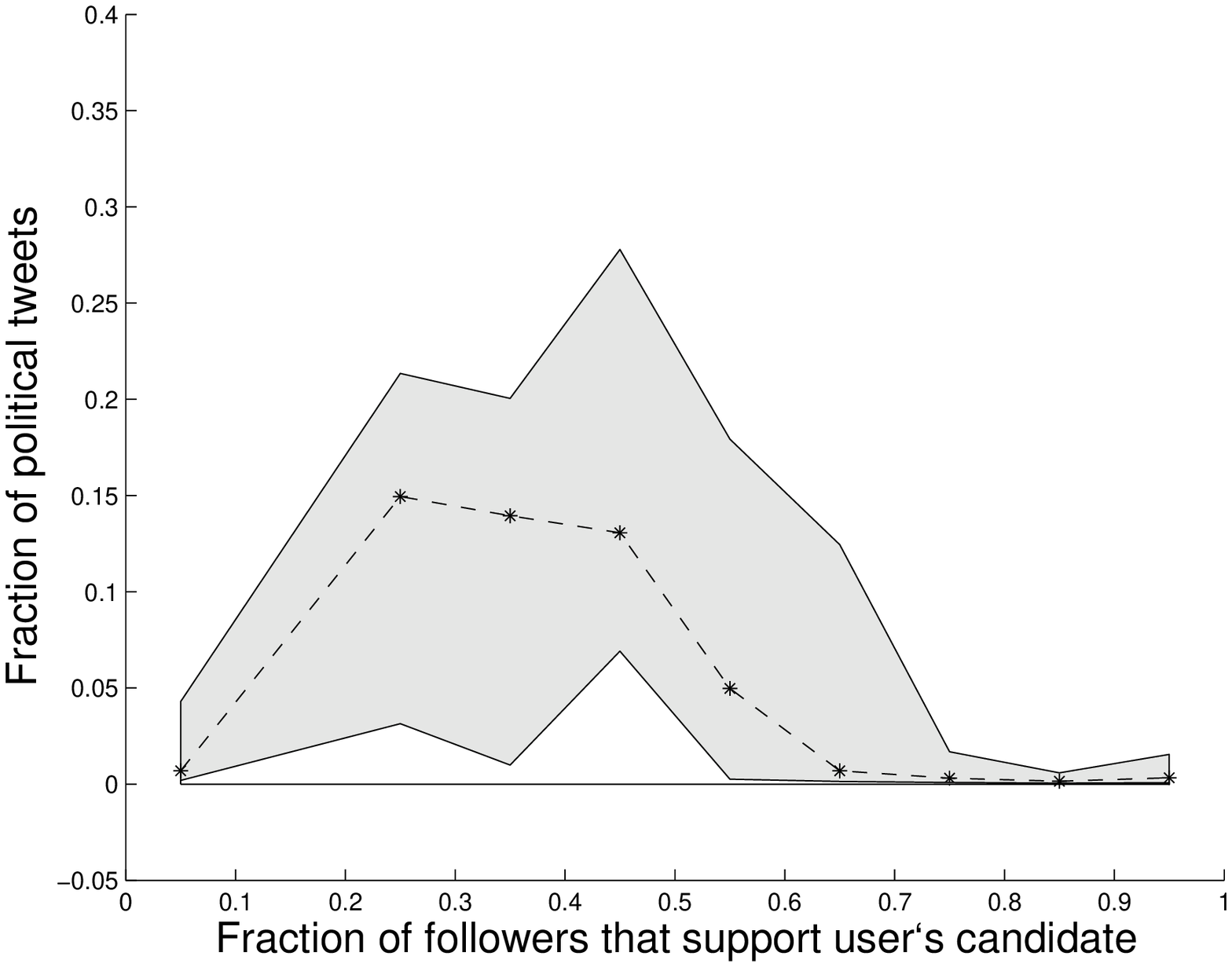}
\caption{Median PA (fraction of political tweets) vs. follower-LM (fraction of users' followers that support their candidate)
for pro-Obama users. Median PA is depicted as a dashed line.
Lower confidence interval is given by the 20th percentile and the upper confidence interval
is given by the 80th percentile. Each bin contains at least $20$ points. One bin with less than $20$ points was omitted.}\label{fig:follower_LM_Dems}
\end{minipage}\hfill
\begin{minipage}{0.45\textwidth}
\centering
\includegraphics[scale=0.4]{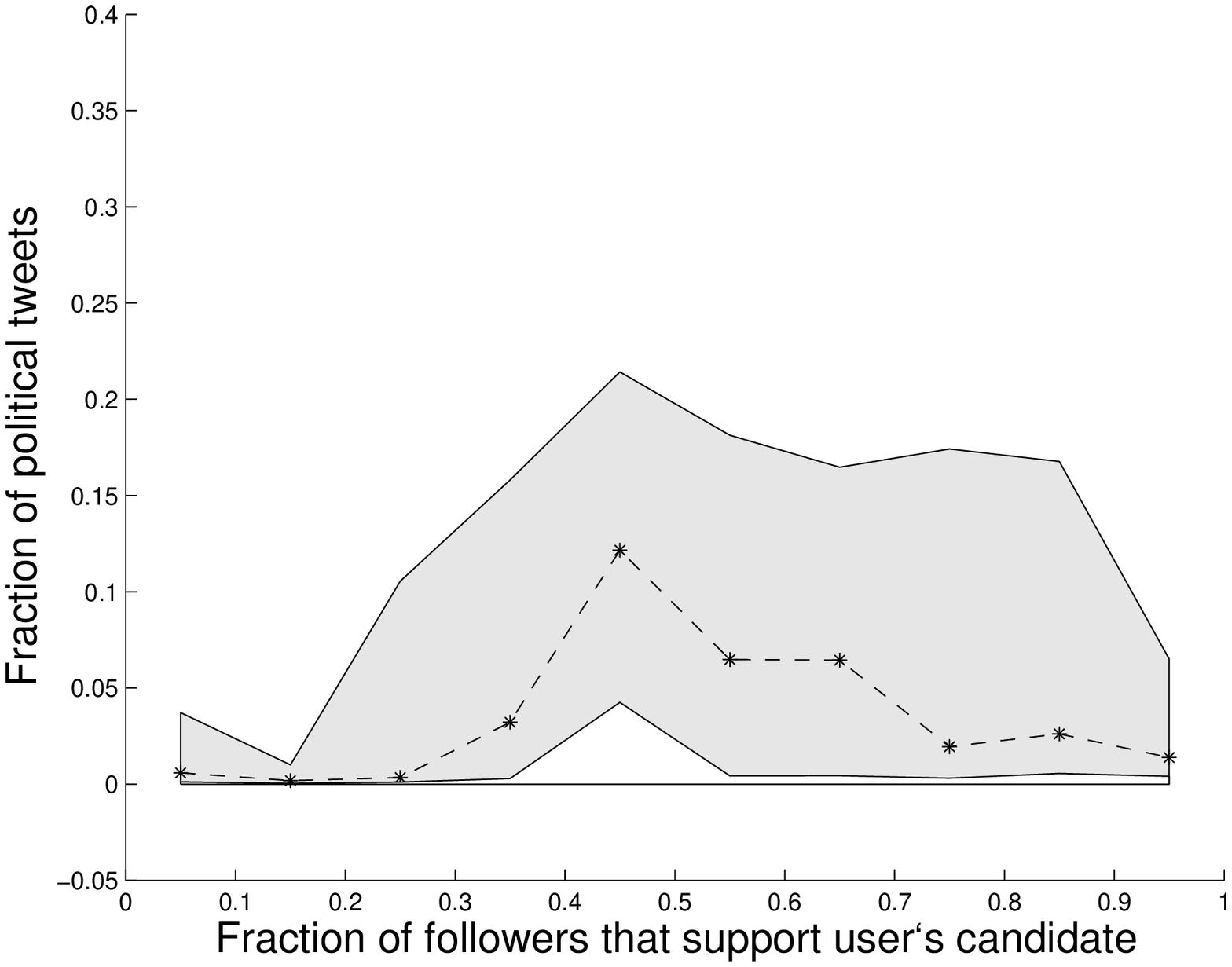}
\caption{Median PA (fraction of political tweets) vs. follower-LM (fraction of users' followers that support their candidate)
for pro-Romney users. Median PA is depicted as a dashed line.
Lower confidence interval is given by the 20th percentile and the upper confidence interval
is given by the 80th percentile. Each bin contains at least $20$ points.}\label{fig:follower_LM_Reps}
\end{minipage}
\begin{minipage}{0.4\textwidth}
\centering
\includegraphics[scale=0.4]{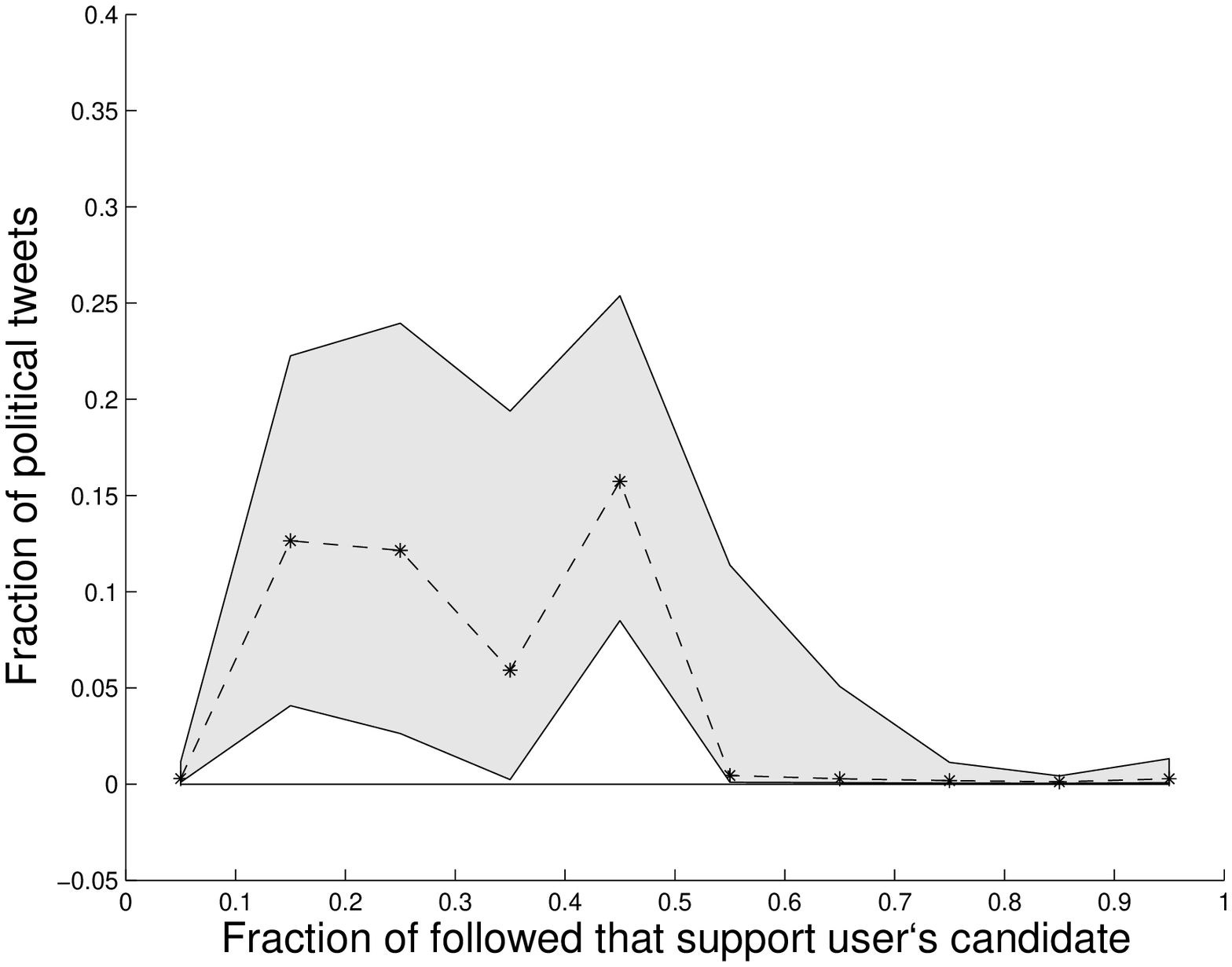}
\caption{Median PA (fraction of political tweets) vs. followed-LM (fraction of people followed by the user that support user's candidate)
for pro-Obama users. Median PA is depicted as a dashed line.
Lower confidence interval is given by the 20th percentile and the upper confidence interval
is given by the 80th percentile. Each bin contains at least $20$ points.}\label{fig:followed_LM_Dems}
\end{minipage}\hfill
\begin{minipage}{0.45\textwidth}
\centering
\includegraphics[scale=0.4]{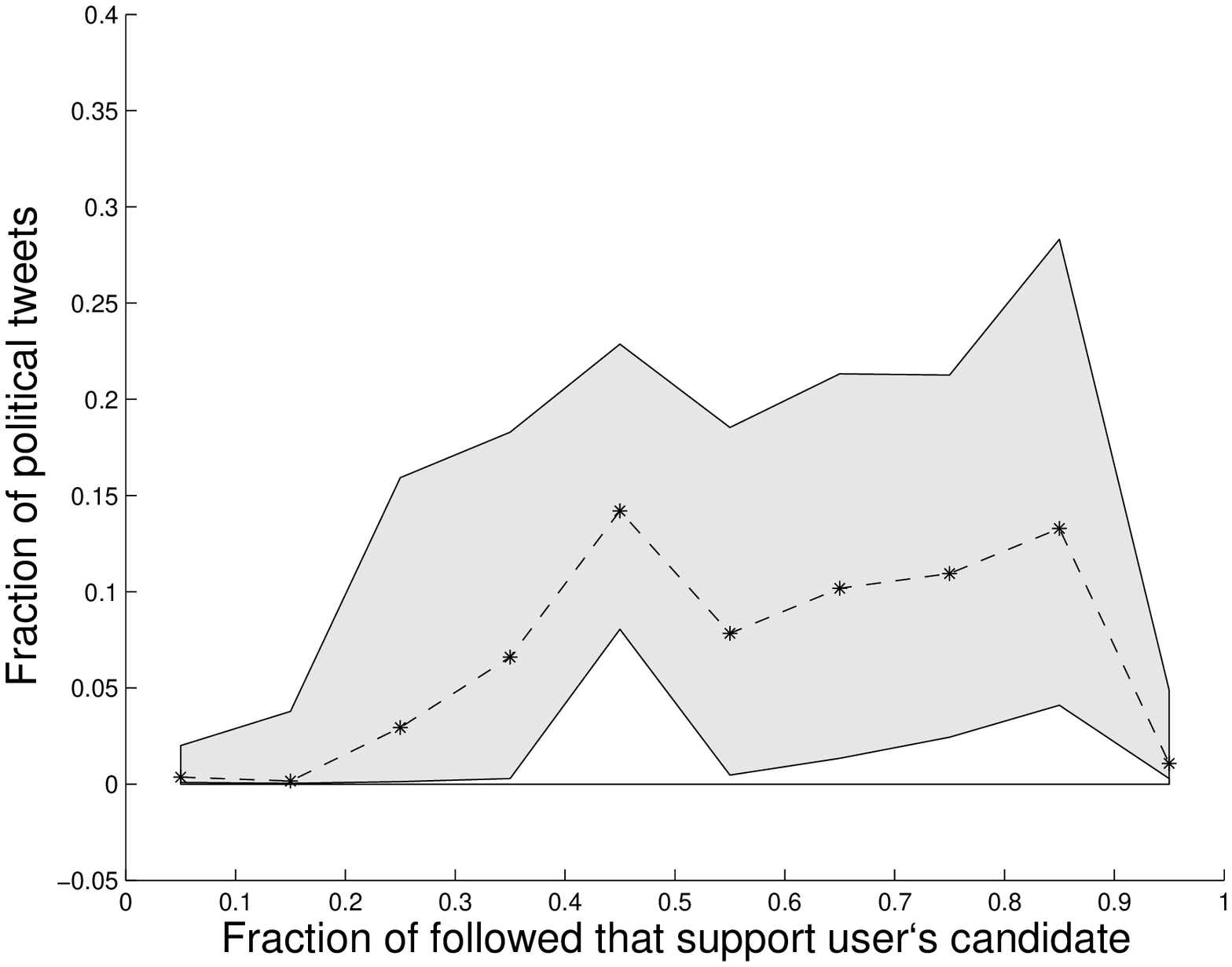}
\caption{Median PA (fraction of political tweets) vs. followed-LM (fraction of people followed by the user that support user's candidate)
for pro-Romney users. Median PA is depicted as a dashed line.
Lower confidence interval is given by the 20th percentile and the upper confidence interval
is given by the 80th percentile. Each bin contains at least $20$ points.}\label{fig:followed_LM_Reps}
\end{minipage}
\begin{minipage}{0.4\textwidth}
\centering
\includegraphics[scale=0.4]{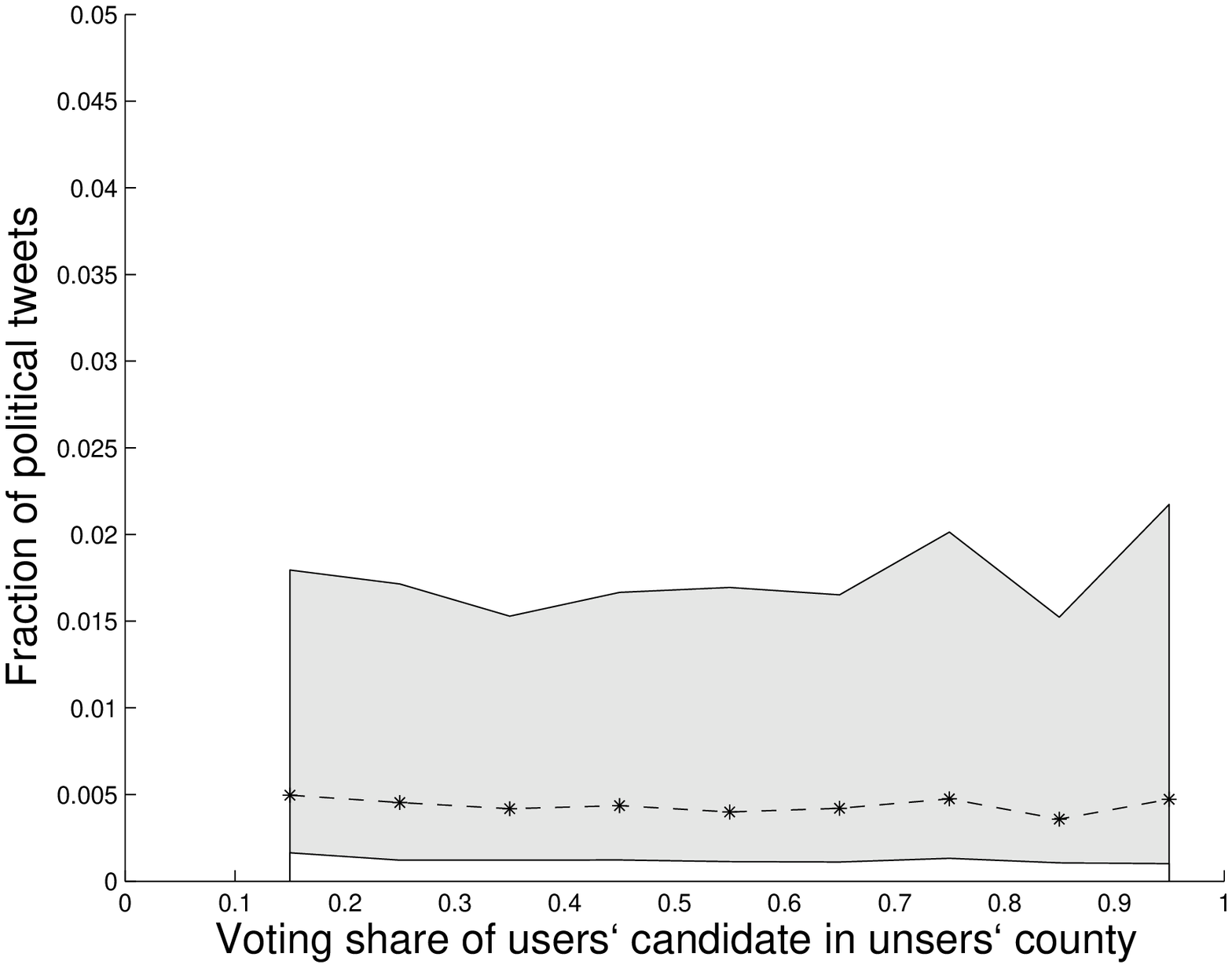}
\caption{Median PA (fraction of political tweets) vs. geographical-LM
(voting share users' candidate in their county).
Median PA is depicted as a dashed line.
Lower confidence interval is given by the 20th percentile and the upper confidence interval
is given by the 80th percentile. Each bin has at least 10 points. Three bins with less than $10$ points were omitted.}\label{fig:geoLM_Dems}
\end{minipage}\hfill
\begin{minipage}{0.45\textwidth}
\centering
\includegraphics[scale=0.4]{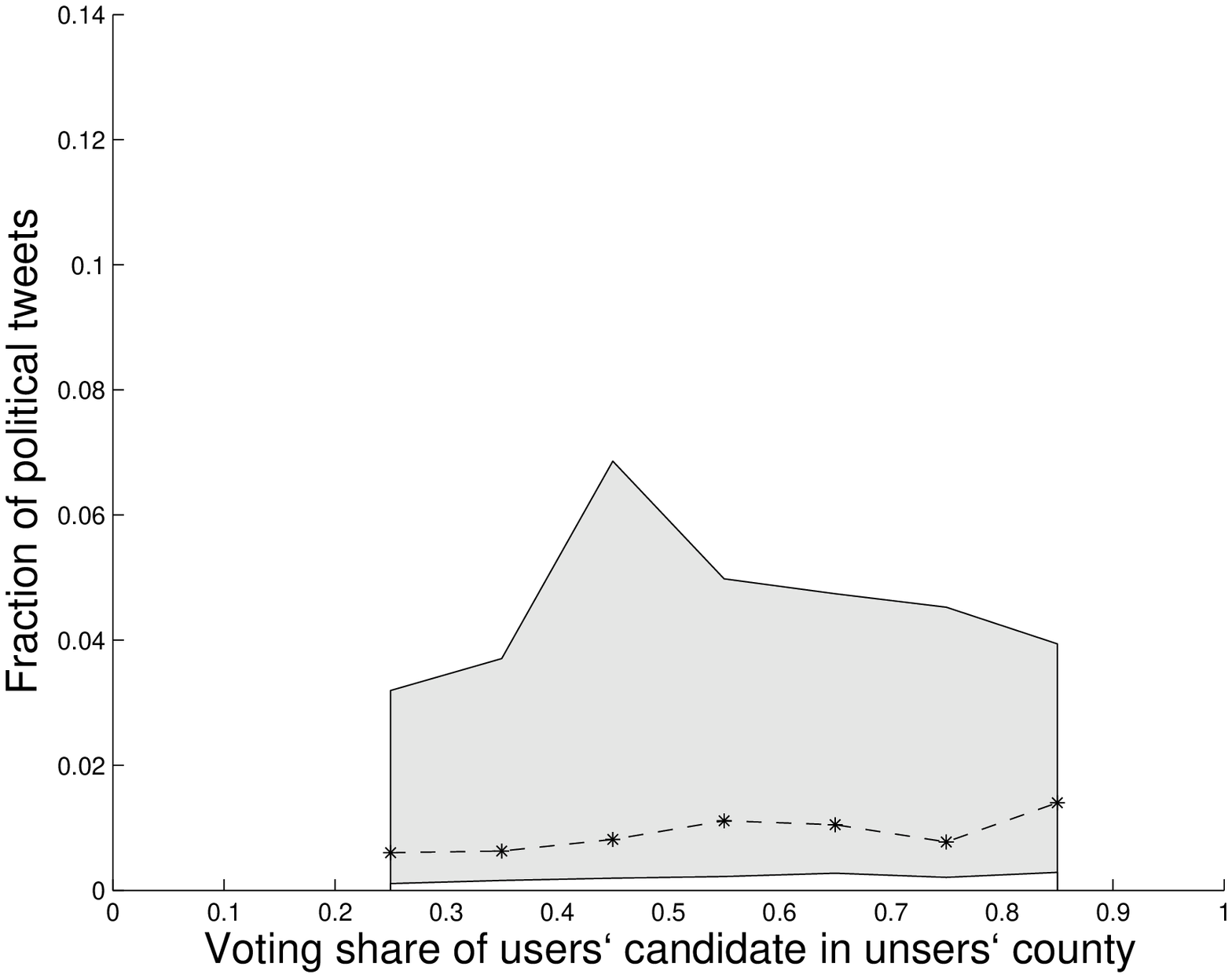}
\caption{Median PA (fraction of political tweets) vs. geographical-LM
(voting share users' candidate in their county). Median PA is depicted as a dashed line.
Lower confidence interval is given by the 20th percentile and the upper confidence interval
is given by the 80th percentile. Each bin has at least 10 points. Three bins with less than $10$ points were omitted.}\label{fig:geoLM_Reps}
\end{minipage}
\end{figure}

We begin by using data from Follower-DS to analyze the dependence of the median level of political activity
on the likemindedness of users that read posts of the considered user. I.e., the dependence of median PA on follower-LM.
To this end, we divide the range $[0,1]$ of possible values of follower-LM to $10$ equally-sized bins. For each bin, we calculate the median PA of
all users with the value of follower-LM that falls in this bin.
Figure \ref{fig:follower_LM_Dems} and Figure \ref{fig:follower_LM_Reps} depict these dependencies for
pro-Obama and pro-Romney users, respectively.

Both for pro-Obama and pro-Romney users we observe a unimodal dependency of median PA on follower-LM.
The single peak corresponds to the disagreement effect and is
obtained for medium values of follower-LM (between $0.2$ and $0.6$ for pro-Obama users and
between $0.4$ and $0.7$ for pro-Romney users).
The level of political activity of politically isolated users (follower-LM smaller than $0.2$ for pro-Obama users
and smaller than $0.4$ for pro-Romney users) is negligible. This observation is also in line with the disagreement effect.
The level of political activity of users in the echo chamber environment (large values of follower-LM) is also very low,
which is in contradiction to the echo chamber amplification theory.

%
%

We now use data from Followed-DS to analyze the dependence of the median level of political activity
on the likemindedness of the Twitter feeds read by the considered user, i.e. the dependence of
median PA on followed-LM.
Again, we divide the range $[0,1]$ of possible values of followed-LM to $10$ equally-sized bins.
For each bin, we calculate the median PA of all users with the value of followed-LM that falls in this bin.
Figure \ref{fig:followed_LM_Dems} and Figure \ref{fig:followed_LM_Reps} depict these dependencies for
pro-Obama and pro-Romney users, respectively.
For both pro-Obama and pro-Romney users, small values of followed-LM (below $0.1$) and high values of followed-LM (above $0.9$)
correspond to very low levels of political activity. The first observation is in line with the disagreement effect and
the second observation contradicts the echo chamber amplification theory. For both pro-Obama and pro-Romney users,
the dominant peak corresponds to the disagreement effect and is obtained for values of followed-LM close to $0.4$.
However, in contrast to the dependence of PA on follower-LM, there are also secondary peaks.
For Obama supporters, there is a secondary peak for the values of followed-LM between $0.1$ and $0.3$.
We hypothesize that this secondary peak is also a manifestation of the disagreement effect.
For Romney supporters, there is a secondary peak for the values of followed-LM between $0.6$ and $0.9$.
This secondary peak may be a manifestation of both the disagreement effect and the echo chamber amplification.

Finally, we analyzed the relationship between the median PA and geographical-LM.
Similarly to figures above, we divide the range $[0,1]$ of possible values of geographical-LM to $10$
equally-sized bins. For each bin, we calculate the median PA of all users with the value of geographical-LM
that falls in this bin.
The results are depicted on Figure \ref{fig:geoLM_Dems} for pro-Obama users and on Figure \ref{fig:geoLM_Reps} for pro-Romney users.
In contrast to the dependence of median PA on likemindedness of user's virtual environment (follower-LM and followed-LM),
the dependence of PA on geographical-LM does not differ for pro-Obama and pro-Romney users.
In fact, it seems that the level of political activity of users is independent of geographical-LM, hence
does not exhibit neither the echo chamber amplification nor the disagreement effect.

\onecolumn
\twocolumn

\section{Discussion}
In this paper we analyzed the connection between user's level of political activity on Twitter (PA)
and the likemindedness of its virtual (follower-LM, followed-LM) and geographical environments (geographical-LM).
Specifically, we focused on the presence of the echo chamber amplification and the disagreement effect.

We showed that user's PA as a function of follower-LM has a similar form for both
pro-Obama and pro-Romney users. In both cases,
high level of political activity of a user correlates with a politically diverse set of followers,
i.e., readers of user's Twitter feed. The dependence of PA on follower-LM exhibits strong disagreement effect,
but does not manifest the echo chamber amplification.

The dependence of user's PA on followed-LM is different for users supporting different candidates.
Pro-Romney users exhibit high level of political activity when Twitter feeds they follow are predominantly,
but not purely, pro-Romney. Namely, pro-Romney users are most active in the mostly likeminded environments.
This behavior is a manifestation of the combination of the disagreement effect and the echo chamber amplification.
In contrast, pro-Obama users tend to have a high level of political activity in the politically adverse environments.
Namely, when Twitter feeds they follow are predominantly pro-Romney but not exclusively pro-Romney.
This behavior is in line with the disagreement effect and contradicts the echo chamber amplification theory.

We also show that the level of users' political activity is independent of
the likemindedness of their geographical environment, for both pro-Obama and
pro-Romney users. In particular, this implies that the dependence of PA on
geographical-LM does not manifest neither the disagreement effect nor the echo chamber amplification.

We thus conclude that the level of political activity of the Twitter users correlated with
likemindedness of their virtual environment and is independent of the likemindedness of their geographical environment.
The exact form of correlation between the PA and the likemindedness of the virtual environment
is in line with the disagreement effect and,
with the exception of PA as a function of followed-LM for Romney supporters,
contradicts the echo chamber amplification theory.

The main limitation of our approach is in the selection of users for our analysis. Specifically,
we ignored users that are politically active but did not express explicit support for neither of
the presidential candidates of the $2012$ US presidential election.
This obviously introduced a bias to our measurements of user's follower-LM and followed-LM.

\bibliographystyle{abbrv}
\bibliography{bibliography}

\end{document}